
\magnification=1200
\overfullrule=0pt
\baselineskip=19pt

\def\ha{Heisenberg}
\def\mft{mean-field theory}
\def\nn{nearest-neighbor}
\def\gs{ground state}
\def\gss{ground states}
\def\cl{correlation}
\def\fr{frustrated}

\def\co{colinear}
\def\sw{spin-wave}
\def\b{boson}
\def\fig{configuration}
\def\v{variational}
\def\e{equations}
\def\gs{ground state}
\def\zte{temperature}

\def\exi{$X_{\vec \delta}$}
\def\yi{$Y_{\vec \delta}$}
\def\mi{${\mu}_{\vec k}$}
\def\ni{${\nu}_{\vec k}$}
\def\wi{${\omega}_{\vec k}$}

\hfill{IISc-CTS-93-7}

\hfill{IP/BBSR/93-58}

\hfill{September, 1993}

\centerline{\bf A bosonic mean-field theory for frustrated Heisenberg}
\centerline{\bf antiferromagnets in two dimensions}

\vskip .2in

\centerline{R. Chitra\footnote*{E-mail: chitra@cts.iisc.ernet.in}}
\centerline{\it Physics Department, Indian Institute of Science,}
\centerline{\it Bangalore 560012, India}

\vskip .1in
\centerline{Sumathi Rao\footnote{**}{E-mail: sumathi@iopb.ernet.in}}
\centerline{\it Institute of Physics, Sachivalaya Marg,}
\centerline{\it Bhubaneswar 751005, India}

\vskip .1in
\centerline{and}

\vskip .1in
\centerline{Diptiman Sen\footnote{***}{E-mail: diptiman@cts.iisc.ernet.in}}
\centerline{\it Centre for Theoretical Studies, Indian Institute of Science,}
\centerline{\it Bangalore 560012, India}

\vskip .4in

\line{\bf Abstract}

\vskip .2in

We use a recently developed bosonic mean-field theory (MFT) to
study the ordered ground states of frustrated Heisenberg
antiferromagnets (FHAFM) in two dimensions. We emphasize the
role of condensates in satisfying the MF variational equations
and their relation to spin correlation functions at low
temperatures. Our results are similar to those obtained using
Schwinger boson MFT. However, we emphasize here that our MFT has
three bosons at each site and that it is not necessary to rotate
the quantization axes on appropriate sub-lattices to align all
the spins ferromagnetically.  This MFT is also closely related
to a new spin-wave theory which enables us to obtain the
spin-wave spectrum easily (without any rotation of axes) for an
entire class of three-dimensionally ordered states. For the
FHAFM on a triangular lattice, we use this theory to compute the
spin-wave spectrum for all values of $~J_2 ~/J_1 ~$ and
demonstrate the phenomenon of order from disorder.

\vskip .2in

\line{PACS numbers: ~75.10.Jm, ~75.50.Ee, ~67.40.Db \hfill}

\vfill
\eject

\line{\bf I. ~INTRODUCTION \hfill}

\vskip .1in

The ground state and low-energy excitations of  \ha\
antiferromagnets (AFM) in two dimensions have been extensively
studied [~1 - 15~], particularly after the discovery of
high-$T_c$ superconductors. The various techniques used for this
purpose include non-linear field theories [~2 - 4~], spin-wave
analysis [~5~], bosonic and fermionic \mft\  [~6 - 13~] and
numerical methods [~14, 15~]. For unfrustrated AFMs such as \ha\
models on bipartite lattices with just \nn\ couplings, it is now
well-established that the zero temperature \gs\ is ordered for
all values of the spin $S$. At any finite temperature $T$, the
models are disordered with a \cl\ length which grows
exponentially with $~1/T~$ for small $T$. However for {\it
frustrated} AFMs (e.g. \ha\ models on bipartite lattices with
next \nn\ interactions as well), the situation remains
interesting and unresolved. Besides the conventionally ordered
states expected by extrapolation from the classical limit, many
other interesting \gs s , such as spin nematics [~5~], flux
phases [~8~] and chiral spin-liquids [~9~] have been proposed.
The issue of which of these states is actually a \gs\ for
different ranges of parameters (i.e. the phase diagram) of
\fr\ spin models remains open.

In this paper, we do not address the question of the complete
phase diagram of \fr\ models. Instead, we concentrate on
studying the properties of the helicoidal (ordered) phases that
are obtained by using a \b ic MF approach. Our main aim is to
further develop a recently introduced \b ic MFT [~16~] and show
how various properties of the MF state can be easily obtained.
We also compare our MFT, which has {\it three}  \b s at every
site (3BR), to the more conventionally used Schwinger boson MFT
which uses two \b s at each site (2BR), and explain both the
merits and the limitations of our approach. Our 3BR has the
further advantage that it is closely linked with a new \sw\
theory, which can be applied even when the spin ordering is not
along any line or plane, but is fully three-dimensional.  We
explore this connection in this paper and also use the \sw\
theory to find the spectrum and demonstrate order from
disorder, in the most general case, for the FHAFM  on a
triangular lattice.

The rest of the paper is organised as follows. In Sec. II, we
study simple examples of the kinds of classical ground states
possible for \fr\ models. This is mainly a review of the models
considered by Villain [~17~]. We concentrate on commonly
occuring examples.  In particular, we study four categories: (i)
the planar spirals, (ii) the colinear spirals, (iii) a
one-parameter family of degenerate canted states, and (iv) a
two-parameter family of degenerate three-dimensional canted
states.  Using our new \sw\ formalism (inspired by the 3BR), we
work out the general \sw\ spectrum in all four categories. In
categories (iii) and (iv), we find that quantum fluctuations
always break the classical degeneracy so that the \sw\ analysis
picks a finite set of \gss\ as having the lowest energy. This
phenomenon is called order from disorder [~18~].  In all the
cases that we have examined so far, the states with the minimum
quantum energy always turn out to be \co\ spirals. As a
particular example, we present the complete calculation of the
\sw\ spectrum and the phenomenon of order from disorder for a
\fr\ model on a triangular lattice. A previous analysis [~18~]
of this model did not include all the type (iv) states.

In Sec. III, we introduce the 3BR and develop the MF variational
approach. The 3BR uses a representation of the spins by three \b
s transforming under the adjoint representation of the $SO(3)$
group. A simple MF picture then leads to certain variational
equations.  At zero temperature, we show that condensates of the
three \b s at the Goldstone modes ($i.e.$, zeros of the \sw\
energy) are required to satisfy these equations. These
condensates are related to long range order (LRO) in the system.
However, at any finite temperature, LRO breaks down (the
condensates vanish) and the spectrum develops gaps proportional
to $~1/T~$ for low temperatures. This is studied in Sec. IV.  An
interesting feature of this MFT is its intimate connection to
the \sw\ theory introduced in Sec. II. We explore this
connection and show that for $S ~\to ~\infty~$, the two
calculations are identical, except that the 3BR produces a
spurious tripling of the \sw\ modes.  This multiplicity is an
unfortunate feature of all \b ic MFTs. The commonly used 2BR
also produces a spurious doubling of modes. We end the paper
with a brief discussion in Sec. V, where we briefly point out
directions for future research.

\vskip .2in

\line {\bf II. ~CLASSICAL GROUND STATES AND SPIN-WAVE THEORY \hfill}

\vskip .1in

Let us consider a general Bravais lattice in $d$ dimensions and
a \ha\ spin Hamiltonian of the form
$$H ~=~ {1\over 2} ~\sum_{\vec n, \vec {\delta}} ~J_{\vec \delta}~
{\vec S}_{\vec n} \cdot {\vec S}_{\vec n + {\vec \delta}} .
\eqno(2.1)$$
Here the sum over $\vec n$ runs over the positions of all the
sites and $\vec \delta$ denotes the position of a neighboring
site.  We assume that the couplings $~J_{\vec \delta} \geq 0~$
for all $\vec \delta$ and that $~J_{\vec \delta}~$ goes to zero
sufficiently rapidly for large $\vec \delta$. In part (A) of
this Section, we tabulate the different kinds of classical \gs\
of this model on square and triangular lattices, for various
ratios of the \nn\ and next \nn\ couplings. In part (B), we
introduce a new Holstein-Primakoff representation of spins to
obtain the spectrum in all the cases tabulated in part (A).
Finally, in part (C), we explicitly compute the dispersion for
the \fr\ model on a triangular lattice. We show how the
zero-point fluctuations always pick out the \co\ states as the
lowest energy states even though, classically, planar canted and
three-dimensional canted states are allowed as \gs s.

\vskip .2in

\line{\bf A. ~Classical Ground States \hfill}

\vskip .1in

The classical \gs s of the Hamiltonian in Eq. (2.1) have been
studied in detail in Ref. [~17~]. Here we look at examples relevant
to \fr\ models on square and triangular lattices.

\noindent
Category (i): In general, the classical \gss\
have the spin configuration
$$\langle~{\vec S}_{\vec n}~
\rangle ~=~ S ( ~\vec {u_1} ~\cos {{\vec Q} \cdot
{\vec n}} ~+~ \vec {u_2} ~\sin {{\vec Q} \cdot {\vec n}}~)
\eqno(2.2)$$
where $\vec {u_1}$ and $\vec {u_2}$ are two orthonormal vectors
defining a plane, and the pitch vector $\vec Q$ is obtained by
minimizing the classical energy of the Hamiltonian given by
$${{E_0 (\vec Q)} \over {NS^2}} ~=~ {1 \over 2} ~\sum_{\vec
\delta} ~ J_{\vec \delta} ~\cos {\vec Q} \cdot {\vec \delta}
\eqno(2.3)$$
where $N$ is the total number of sites. This configuration is called
the planar spiral configuration.

\noindent
Category (ii):
Let us use $\vec \tau$ as a generic symbol to denote a
reciprocal lattice vector so that ${\vec \tau} \cdot \vec n ~=~
2 \pi~$ for all sites $\vec n$. Then for the special case $~2~
\vec Q ~=~ {\vec \tau}~$, the configuration in Eq. (2.2) reduces
to
$$\langle~{\vec S}_{\vec n}~\rangle ~=~ S ~\vec {u_1} ~\cos
\vec Q \cdot {\vec n} ~=~ \pm ~S ~\vec {u_1}
\eqno(2.4)$$
which describes a \co\ spiral.

\noindent
Category (iii): When the minimum of $~E_0 (\vec Q)~$ in Eq.
(2.3) occurs for {\it two} different vectors ${\vec Q}_1~$ and
${\vec Q}_2~$, such that $~2 {\vec Q}_1~=~ {\vec \tau}_1 ~$and
$~2 {\vec Q}_2~=~ {\vec \tau}_2~ $ but ${\vec Q}_1+ {\vec Q}_2
\ne {\vec \tau}_3 ~$, then there exists a one-parameter family
of configurations which have minimum energy, given by
$$\eqalign{\langle~~{\vec S}_{\vec n} ~\rangle &=~ S ~(~{\vec u}_1 ~\cos
{\vec Q}_1 \cdot \vec n~ \cos {\phi} ~+~
{\vec u}_2 ~{\cos {{\vec Q}_2 \cdot \vec n}} ~\sin{\phi}~) \cr
&=~ S ~(~\pm ~{\vec u}_1 ~\cos {\phi} ~\pm ~{\vec u}_2 ~\sin {\phi} ~). \cr}
\eqno(2.5)$$
These configurations describe planar canted states.

\noindent
Category (iv): When the minimum of $E_0(\vec Q)$ occurs for {\it three}
different vectors ${\vec Q}_1, {\vec Q}_2 ~$ and ${\vec Q}_3 ~$, all of which
satisfy $~2 {\vec Q}_i ~=~ \vec \tau_i ~$, then there exists a two-parameter
family of \gs\ configurations given by
$$\eqalign{\langle ~{\vec S}_{\vec n} ~\rangle &=~
S ~({\vec u}_1  ~\cos {{\vec Q}_1
\cdot \vec n} ~\sin {\theta} ~\cos {\phi} ~+~ {\vec u}_2 ~\cos {{\vec Q}_2
\cdot \vec n} ~\sin {\theta} ~\sin{\phi} ~+ ~{\vec u}_3 ~\cos {{\vec {Q_3}}
\cdot \vec n} ~\cos {\theta}) \cr
&=~ S ~(~\pm ~\vec {u_1} ~\sin {\theta} ~\cos{\phi} ~\pm ~\vec {u_2} ~
\sin{\theta}~ \sin{\phi} ~\pm ~\vec {u_3} ~\cos{\theta} ~) \cr}
\eqno(2.6)$$
where $(\vec {u_1}, \vec {u_2}, \vec {u_3}) ~$ form an
orthonormal triad and $~0 ~\le ~\theta ~\le ~\pi~$, $~0 ~\le
{}~\phi ~\le~ \pi / 2 ~$.  These configurations describe an
ordered state in three dimensions and are called
three-dimensional canted states. Notice that the planar canted
states in (iii) are a special case of these canted states for
$~\theta ~=~\pi/ 2~$ or for $~\phi ~=~ 0~$ or $~\pi / 2$. The
\co\ spiral in (ii) is, in turn, a special case of the planar
canted state for $\phi~=~ 0~$ or $~ \pi / 2$.

All these kinds of \gss\ occur for the $~J_1 ~-~ J_2~$ models on
square and triangular lattices. ($J_1 ~$ is the \nn\ coupling
and $J_2~$ is the next \nn\ coupling). In Table I, we show the
different kinds of classical \gs s that occur for different
ratios of the coupling constants.

\vskip .2in

\line{\bf B. ~Spin-Wave Spectrum \hfill}

\vskip .1in

We now introduce a variant of the Holstein-Primakoff (HP)
representation of the spins to obtain the \sw\ spectrum. This
formalism can be used to expand about any classical \fig\ as
shown below.  Consider a spin with classical components given by
$$\eqalign{\langle ~S_z ~\rangle &=~ S ~\cos{\alpha}, \cr
\langle ~S_x ~\pm ~i S_y ~\rangle &=~ S ~e^{\pm i \beta} ~\sin
{\alpha}. \cr} \eqno(2.7)$$
Bosonic operators $~b~$ and $~b^{\dag}~$ are introduced ({\it
{$\grave a$} la} HP) to represent deviations away from the
classical \fig\ by
$$\eqalign{S_z =~ \cos{\alpha} ~&(S ~-~ b^{\dag} ~b) ~-~ {1 \over 2} ~
\sin {\alpha}~ [~(2S ~-~ b^{\dag} ~b)^{1 / 2} ~b ~+~ h.c. ~]~, \cr
S_x ~\pm ~i S_y =~ e^{\pm i \beta} ~&[ ~\sin{\alpha} ~(S ~- ~b^{\dag} ~b)~
+ ~{1 \over 2} ~\cos{\alpha} ~\lbrace ~( 2S ~-~
b^{\dag} ~b)^{1 / 2}b ~+ ~h.c.~ \rbrace \cr
&\pm ~{1\over 2} ~ \lbrace ~(2S ~- ~b^{\dag} ~b)^{1/2} ~~b ~-~ h.c.~
\rbrace ~]. \cr}
\eqno(2.8)$$
It can be easily checked that these spin operators satisfy the
usual commutation relations if $~[ ~b, ~b^{\dag} ~]~=~1 $.
Notice that the usual HP \b s are recovered if we set
$\alpha ~=~ 0~$ or $~\pi~$.

To obtain the \sw\ spectrum, we expand Eq. (2.8) to next-to-leading
order in $1/S$. We then find that
$$\vec S ~=~ \langle ~\vec S ~\rangle \bigl(1 ~-~ {{b^{\dag} ~b }\over S}~
\bigr)~
- ~b ~\langle ~\vec a ~{\rangle}^\ast ~- ~b^{\dag} ~\langle ~\u
c a ~\rangle
\eqno(2.9)$$
where $\langle ~\vec a ~\rangle$ denotes the vector with components given by
$$\eqalign{\langle ~a_1 ~\rangle &=~ \sqrt {S \over 2} ~(~ - ~\cos {\alpha}~
\cos{\beta} ~+~ i ~\sin{\beta}~) \cr
\langle ~a_2 ~\rangle &=~ {\sqrt {S \over 2}} ~(~- ~\cos \alpha ~\sin \beta~
- ~i ~\cos \beta~) \cr
\langle ~a_3 ~\rangle &=~ {\sqrt{S \over 2}} ~\sin \alpha. \cr}
\eqno(2.10)$$

\noindent
The reason for introducing the vector $\vec a$ will become clear in
Sec. III, where
the three $a_i$'s will be identified with three bosonic operators
and $\langle ~a_i ~\rangle$ will denote the \gs\ expectation values of
these \b s. Here, $\vec a$ is just a convenient notation.
Notice that $~{\langle ~\vec S ~\rangle } / S~$ , $~\sqrt{2 / S} ~Re~
\langle ~\vec a ~\rangle~ $ and $~\sqrt{2 / S} ~Im~ \langle ~\vec a ~
\rangle ~$ form an orthonormal triad so that
$$\langle ~\vec S ~\rangle \cdot ~\langle~\vec a ~
\rangle ~=~ \langle~ \vec S ~\rangle \cdot {\langle~ \vec a ~\rangle }^{\ast} ~
=~0.
\eqno(2.11)$$

In part (A), we classified states of the spin model where
$\langle ~{\vec S}_{\vec n} ~\rangle \cdot \langle ~{\vec S}_{\vec n ~+ ~{\vec
\delta}} ~\rangle~ $ is independent of $\vec n$ and only depends on
$\vec \delta$ through the
angle $~\phi_{\vec \delta}~$ between the classical spins at $\vec n$ and
$\vec n + \vec \delta~$ -~$i.e.$,
$$\langle~ {\vec S}_{\vec n} ~\rangle \cdot
\langle~ {\vec S}_{\vec n ~+~ {\vec \delta}} ~\rangle ~=~ S^2 ~\cos
{\phi_{\vec \delta}}.
\eqno(2.12)$$
In this part, we make the further assumption that
$$\eqalign{X_{\vec \delta}~ &=~ \langle~ {\vec a}_{\vec n} ~\rangle \cdot
\langle~ {\vec a}_{\vec {n} ~+~ \vec \delta} ~\rangle \cr
Y_{\vec \delta}~ &=~ \langle~ {\vec a}_{\vec n} ~\rangle \cdot \langle~
{\vec a}_{\vec {n} ~+ ~\vec \delta} ~\rangle^{\ast} \cr}
\eqno(2.13)$$
are also independent of $\vec n$ and depend only on $\vec \delta$.
(We shall see later that this assumption is always valid in
categories (i) - (iii), but in (iv), it holds only if there exists
an extra condition relating the vectors $\vec {Q_1}, ~\vec {Q_2}~$
and $\vec {Q_3}~$). $~X_{\vec \delta}~$ (and $~Y_{\vec \delta}~$) are
called the short-range antiferromagnetic
(and ferromagnetic) order parameters respectively. From Eq. (2.10), we get
$$\eqalign{\vert ~X_{\vec \delta}~ \vert &=~ S ~{\sin}^2 (~\phi_{\vec
\delta} / 2~) \cr
{\rm and}\quad
\vert ~Y_{\vec \delta} ~\vert &=~ S ~
{\cos}^2 (~\phi_{\vec \delta} / 2~). \cr}
\eqno(2.14)$$
Thus, their magnitudes equal $~S~$ (and $~0~$) if $\phi_{\vec \delta}~=~ \pi$
and $~0~$ (and $~S~$) if $\phi_{\vec \delta}~=~0$,
justifying their nomenclature.

Finally, let us define
$$\eqalign{\lambda~ &=~ - ~S ~\sum_{\vec \delta} ~J_{\vec \delta}~
\cos {\phi_{\vec \delta}} \cr
{\rm and}\quad
{\vec h}_{\vec n}~ &=~ \sum_{\vec \delta} ~J_{\vec \delta} ~\langle~
{\vec S}_{\vec n ~+~ \vec {\delta}} ~\rangle. \cr}
\eqno(2.15)$$
Here $~{\vec h}_{\vec n}~$ may be interpreted as the local magnetic field at
the site $\vec n$. In the classical \gs\ , $~\langle~ {\vec S}_{\vec n}~
\rangle ~$
at the site $\vec n~$ aligns itself anti-parallel to the local magnetic field
${\vec h}_{\vec n}~$. Thus the orthogonality of $\langle~ \vec S~ \rangle$
and $\langle~ \vec a~ \rangle $ in Eq. (2.10) imply that
$${\vec h}_{\vec n} \cdot \langle~ {\vec a}_{\vec n}~ \rangle ~=~ {\vec
h}_{\vec n} ~\langle~
{\vec a}_{\vec n} ~\rangle^{\ast} ~=~0.
\eqno(2.16)$$
Substituting for ${\vec S}_{\vec n}~$ and ${\vec S}_{\vec n ~+~\vec \delta}$
from Eq. (2.9) in the Hamiltonian in Eq. (2.1) and using Eqs. (2.14 - 2.16),
we find that to $O(S)$ in an expansion in ${1 / S}~$,
$$\eqalign{H ~=~ - ~NS \lambda ~&+~ \lambda ~\sum_{\vec n} ~
b_{\vec n}^{\dag} ~b_{\vec n} \cr
&+~ {1 \over 2} ~\sum_{\vec n , ~{\vec \delta}} ~J_{\vec \delta}~
(~X_{\vec \delta} ~b_{ \vec n ~+~ {\vec \delta}}^{\dag} ~b_{\vec n}^{\dag} ~+~
Y_{\vec \delta} ~b_{\vec n ~+~
{\vec \delta}}^{\dag} ~b_{\vec n} ~+~ h.c. ~). \cr}
\eqno(2.17)$$
The first term in Eq. (2.17) is the classical energy $~E_{cl}~$
of $~O(S^2)~$ and the remaining terms are of $~O(S)$.

To find the spectrum, we transform to momentum space and diagonalize
Eq. (2.17) by a Bogoliubov transformation. This yields
$$H ~=~ -~{{\lambda SN} \over 2} ~+~ N ~\int ~Dk ~{\omega_{\vec k}} ~( ~{\vec
B}_{\vec k}^{\dag} ~{\vec B}_{\vec k} ~+~ {1 \over 2}~)
\eqno(2.18)$$
where the dispersion relation is given by
$$\eqalign{\omega_{\vec k}~ &=~
(\mu_{\vec k}^{2} ~-~\nu_{\vec k}^2)^{1 /2} \cr
{\rm with}\quad\mu_{\vec k}~ &=~ \lambda ~+ ~\sum_{\vec \delta}~
Y_{\vec \delta} ~\cos {\vec k} \cdot \vec \delta \cr
{\rm and}\quad\nu_{\vec k}~ &=~ \sum_{\vec \delta} ~J_{\vec
\delta} ~(~X_{\vec \delta} ~e^{i
\vec k \cdot {\vec \delta}} ~+~ h.c. ~). \cr}
\eqno(2.19)$$
The Bogoliubov transformed bosons are
$$\eqalign{{\vec B}_{\vec k}~ &=~ \cosh \theta ~{\vec b}_{\vec k} ~+~ \sinh
\theta ~{\vec b}^{\dag}_{- \vec k} \cr {\rm and} \quad
{\vec B}^{\dag}_{\vec k}~ &=~ \sinh \theta ~{\vec b}_{- \vec k} ~+~ \cosh
\theta ~{\vec b}^{\dag}_{\vec k} \cr}
\eqno(2.20)$$
and the integration measure is $~Dk~ =~ v ~d^2 k / (2 \pi)^2 ~$
where $~v~$ is the area per unit cell.

The explicit forms for the dispersion in the four categories
defined in part (A) can now be found. For categories (i) - (iii),
$X_{\vec \delta}~$ and $Y_{\vec \delta}~$ as defined by Eq. (2.13) are
independent of $\vec n$ and hence, it suffices to find $\cos {\phi_{\vec
\delta}} ~$. For categories (i) and (ii)
$$\cos {\phi_{\vec \delta}} ~=~ \cos {\vec Q \cdot \vec \delta}.
\eqno(2.21)$$
For category (iii),
$$\cos {\phi_{\vec \delta}} ~=~ \cos {\vec {Q_1} \cdot {\vec \delta}} ~
{\cos}^2 \phi ~+ ~\cos {\vec {Q_2} \cdot {\vec \delta}} ~{\sin}^2 \phi.
\eqno(2.22)$$

\noindent
For category (iv), however, $X_{\vec \delta}~$ and $Y_{\vec \delta}~$ in
Eq. (2.13)
are not independent of $\vec n$, unless ${\cos {\vec {Q_1} \cdot \vec n}}~
{\cos {\vec {Q_2} \cdot \vec n}} ~{\cos {\vec {Q_3} \cdot \vec n}}~=~1~$
for all $\vec n$. Since $~\vec {Q_i}\cdot \vec n ~=~0 ~$, this implies that
$$\vec {Q_1} ~+~\vec {Q_2} ~+~ \vec {Q_3} ~=~ \vec {\tau}.
\eqno(2.23)$$
(Fortunately, this is true for the two examples of category (iv) in
Table I). With this condition satisfied, we have
$${\cos {\phi_{\vec \delta}}} ~=~ {\cos {\vec {Q_1} \cdot {\vec \delta}}} ~
{{\sin}^2 \theta} ~{{\cos}^{2} \phi} ~+~ {\cos {\vec {Q_2}\cdot {\vec
\delta}}}~
{{\sin}^{2} \theta} ~{{\sin}^{2} \phi} ~+~ {\cos {\vec {Q_3} \cdot {\vec
\delta}}} ~{{\cos}^{2} \theta}.
\eqno(2.24)$$
Since $\vec {Q_1}, ~{\vec {Q_2}} ~$ and $\vec {Q_3}~$ are not independent,
Eq. (2.24) may be further simplified. Using Eqs. (2.22) and (2.24),
$~X_{\vec \delta}, ~Y_{\vec \delta}~$ and $\lambda$
and consequently $~\omega_{\vec k}~$ can be computed.

In Table 2, we present the values for the angle $~\phi_{\vec \delta}~$ (given
by Eqs. (2.6) and (2.12)), $~X_{\vec \delta}~$ and $~ Y_{\vec
\delta}~$ for the four different possible values of
$~{\cos {\vec {Q_1} \cdot {\vec \delta}}}~$ and $~{\cos {\vec {Q_2} \cdot
{\vec \delta}}}~$. Thus, we see that this method can be used to find the
\sw\ spectrum about {\it any} classical \gs , even when it has no linear
or planar order, but is fully three-dimensional.

\vskip .2in

\line {\bf C. ~Spin-Wave Spectrum of the $~J_1 ~- ~J_2~$ model on a
Triangular Lattice \hfill}

\vskip .1in

We now apply this new \sw\ theory to the $~J_1 ~- ~J_2~$ model on a triangular
lattice and show that the phenomenon of `order from disorder' occurs.
Although this phenomenon has been demonstrated for the triangular
AFM for some ratios of $~J_2 / J_1 ~$ [~18~], we have been able
to extend those results to the most general case, where the
classical \gs\ has an arbitrary three-dimensional spin ordering.

The sites of a triangular lattice are located at ${\vec n}~=~
a ~[~(n_1 ~+ ~n_2 / 2~) ~{\hat x} ~+ ~{\sqrt {3 / 2}}~ n_2 ~{\hat
y} ~]$, where $~(~n_1~, ~n_2~)~$ are a pair of integers, $a$ is
the lattice spacing and $v~=~ {\sqrt {3 / 2}} ~a^2~$ is the area
per site.  The hexagonal Brillouin zone is shown in Fig. 1.
Define $~\epsilon~=~ J_2 / J_1~$.  For $~\epsilon ~<~ {1/
8}~$, the classical \gs\ is an example of category (i), with the
pitch vector $\vec {Q}$ given by ${\vec Q }~=~(~{4 \pi / 3 a},
0~)~$, denoted by point A in Fig. 1.  Other values of $~\vec Q~$
related to this one by $~\pi / 3~$ rotations in the $~(~k_x~,
k_y~)~$ plane (i.e.  the other five corners of the hexagon) are
also allowed. For $~\epsilon~ >~ 1~$ too, the classical \gs\
falls into category (i) with $~{\vec Q}~=~(~0, ~Q_y~)~$ with
$~\cos (~{\sqrt {3 / 2}} ~a ~Q_y~)~=~ -~(~1 ~ +~ 1 / \epsilon ~)
/ 2 ~$. This is represented by point C in Fig. 1, with other
possible values of $\vec Q$ being related to it by $~\pi / 3~$
rotations.  In both these cases, $~{\cos{\phi_{\vec
\delta}}}~=~{\cos {\vec Q \cdot {\vec \delta}}}~$, $~X_{\vec
\delta}~$ and $~ Y_{\vec \delta}~$ are correspondingly
identified.

However, for $~{1 / 8} ~\leq ~\epsilon ~\leq ~1~$, the classical
\gs\ belongs to category (iv) because $\vec Q$ lies at the
centre of one of the six edges, such as the point B in Fig. 1,
and hence is half of a reciprocal lattice vector. Let us choose
the three vectors as $~{\vec {Q_1}}~=~\pi ~(~1, ~1/ {\sqrt 3} ~)~/a ~$,
${\vec {Q_2}} ~=~ \pi ~(~- 1, ~1 / {\sqrt 3}~ ) ~/a ~$
and ${\vec {Q_3}}~=~ \pi ~(~0, ~-2/ {\sqrt 3} ~)~/a ~$. The
other three possible vectors are equivalent to these upto
reciprocal lattice vectors.  Besides $2 {\vec {Q_i}}~$, we can
check that ${\vec {Q_1}} ~+ ~{\vec {Q_2}} ~+~ {\vec {Q_3}}~$ is
also a reciprocal lattice vector. Hence, $~\phi_{\vec
\delta} ~$, $~X_{\vec \delta}~$ and $~Y_{\vec \delta}~$ can be
found from Table 2. The \sw\ spectrum is given by Eq. (2.19) with
$$\eqalign{{{\mu_{\vec k}} \over {2 J_1 S}} ~=~ &1 ~+~ \epsilon ~+~ {\sin}^2
\theta ~(~{\cos}^2 \phi ~D_1 ~+~ {\sin}^2 \phi ~D_2~) ~-~ {\cos}^2
\theta ~D_3, \cr
{{\nu_{\vec k}} \over {2 J_1 S}} ~=~ &[~ {\sin}^2 \phi ~-~ {\cos}^2 \theta ~
{\cos}^2 \phi ~+ ~i \cos \theta ~\sin 2 \phi~] ~D_1 \cr
&+~ [~{\sin}^2 \phi ~-~ {\cos}^2 \theta ~ {\cos}^2 \phi ~- ~i \cos \theta ~
\sin 2 \phi ~ ]~ D_2 ~+~ {\sin}^2 \theta ~D_3, \cr}
\eqno(2.25)$$
where
$$\eqalign{D_1~ &=~ \cos (~{{k_xa} \over 2} ~-~ {\sqrt {3} \over 2} ~
k_{y}~ a~) ~+~ \epsilon ~\cos (~{3 \over 2} ~k_{x} ~a ~+~ {\sqrt {3} \over
2} ~k_{y} ~a~), \cr
D_2~ &=~ \cos (~{{k_x a} \over 2} ~+~ {{\sqrt 3} \over 2} ~k_{y} ~a~) ~+~
\cos (~{3 \over 2} ~k_{x} ~a ~-~ {{\sqrt 3} \over 2} ~k_{y} ~ a~) \cr
{\rm and} \quad
D_3~ &= ~\cos (~k_{x}~ a~) ~+~ \epsilon ~\cos (~{\sqrt 3} ~k_{y}~ a~). \cr}
\eqno(2.26)$$
Thus, the quantum energy
$${{E_q} \over N}~=~ \int ~{d^2 k \over {(2 \pi)^2 } }~v ~{\omega_{
\vec k} \over 2}~$$
is a function of $~\theta, \phi~$ and $~\epsilon$.
We have numerically studied this for a large number of values
of $~\epsilon~$ from $~1 / 8~$ to $~ 1~$ and $(\theta, ~\phi)$
for $~0 ~\leq ~\theta ~\leq ~{\pi / 2}, ~0 ~\leq~ \phi ~\leq ~{\pi / 2}~$.
Other values of $(\theta, \phi)$ can be related to this range
by rotations and reflections.
(A symmetric tetrahedral spin configuration arises in the special
case $~\theta ~=~{\cos}^{-1} (~1 / {\sqrt 3} ~), ~\phi ~=~\pi / 4~$).
For all $~\epsilon~$, we find that $~E_q~$ is minimum for
the three cases  $~(\theta~=~0~$, any value of $\phi~$) and $~(\theta ~=~
\pi / 2, ~\phi~=~0 ~$ or $~ \pi / 2)~$. These are precisely
the \co\ spirals of category (ii). Thus we have established that
the order from disorder phenomenon occurs for triangular AFMs, where
the general classical \gs\ has three-dimensional ordering.

\vskip .2in

\line{\bf III. ~THREE BOSON REPRESENTATION OF SPINS \hfill}

\vskip .1in

The \sw\ spectrum presented in Sec. II is an expansion about an
ordered classical \gs\ and is only applicable when the system
has LRO.  Moreover, it explicitly breaks rotational invariance
and is only valid in the large-$S$ limit. To overcome these
limitations, different techniques using bosonic and fermionic
operators have been tried.  The most well-studied of these is
the Schwinger boson MFT [~10, 11~]. The Schwinger \b \ MFT
involves a representation of the spins in terms of two \b s
(2BR) with a constraint on the number of \b s at each site.
This representation has the disadvantage that the first step in
the analysis involves a rotation of all the spins to a
`ferromagnetic' alignment. However, a different representation
of the spins was recently introduced [~16~] in terms of three
bosons (3BR) with two constraints at each site. This does not
involve any such rotation and is technically simpler to use,
particularly when the classical \gs\ belongs to the non-colinear
categories (iii) or (iv).

The 3BR represents the spins as
$$\vec S ~=~ - ~i ~{\vec a}^{\dag} \times {\vec a}
\eqno(3.1)$$
where $\vec a$ are three \b ic operators satisfying
$~[~a_i~, ~a_j^{\dag} ~]~=~ {\delta}_{ij} ~$. To satisfy the relation
${\vec S}^2 ~=~ S(S+1)~$, we need to impose two further hermitian
constraints on the bosons given by
$$\eqalign{{\vec a}^{\dag} \cdot {\vec a}~ &=~ S \cr
{\rm and} {\quad} {\vec a}^{\dag} \cdot {\vec a}^{\dag} ~{\vec a} \cdot
{\vec a}~ &=~ 0 \cr}
\eqno(3.2)$$
on all physical states. Since $~{\vec a}^{\dag} \cdot {\vec a}~$ is a
number operator, this clearly indicates that the 3BR only works
for integer values of $S$. In terms of these \b s, the Hamiltonian
in Eq. (2.1) can be written as
$$\eqalign{H~=~ {1 \over 2} ~&\sum_{\vec n, {\vec \delta}}~
J_{\vec \delta} ~[~:~Y_{\vec n, \vec n + {\vec \delta}}^{\dag}~
Y_{\vec n, \vec n + {\vec \delta}} ~: ~-~ X_{\vec n, \vec n + {\vec \delta}}^{
\dag}~ X_{\vec n, \vec n + {\vec \delta}} ~] \cr
+~ &\sum_{\vec n} ~[~ {\lambda}_{\vec n} ~(~{\vec a}_{\vec
n}^{\dag} \cdot ~{\vec a}_{\vec n} ~
-~ S~) ~-~ {\rho}_{\vec n} ~{\vec a}_{\vec n}^{\dag}
\cdot ~{\vec a}_{\vec n}^{\dag} ~{\vec a}_{\vec n}
\cdot ~{\vec a}_{\vec n} ~], \cr}
\eqno(3.3)$$
where we have used the identity
$${\vec S}_{\vec n} \cdot {\vec S}_{\vec m} ~=~ :~Y_{\vec n, \vec m}^{\dag} ~
Y_{\vec n, \vec m}~: ~- ~X_{\vec n, \vec m}^{\dag}~ X_{\vec n, \vec m}
\eqno(3.4)$$
with
$$\eqalign{Y_{\vec n, \vec m}~ &=~ {\vec a}_{\vec n}^{\dag}
\cdot ~{\vec a}_{\vec m} \cr
{\rm and} {\quad} X_{\vec n, \vec m}~ &=~ {\vec a}_{\vec n}
\cdot ~{\vec a}_{\vec m}. \cr}
\eqno(3.5)$$
The colons indicate normal ordering, and $~\lambda_{\vec n}~$ and $~
\rho_{\vec n}~$
are Lagrange multiplier fields introduced to enforce the constraints.

\vskip .2in

\noindent
A. ~Hartree-Fock Mean-Field Theory

\vskip .1in

We make a MF decomposition of the Hamiltonian by writing
$$A^{\dag}~A ~=~ \langle~ A^{\dag} ~\rangle ~A ~+~ A^{\dag} ~\langle~ A~
\rangle~ -~\langle~ A^{\dag} ~\rangle ~\langle~ A~ \rangle
\eqno(3.6)$$
where $A~=~ X_{\vec n, \vec n +{\vec \delta}} ~$ or
$Y_{\vec n, \vec n + {\vec \delta}} ~$. (Such a decomposition
can be justified in the large-$N$ limit, by generalizing the spin
Hamiltonian from $SO(3)$ to $SO(N)~$ [~16~].) Next, we make the MF
ansatz that
$\langle~ X_{\vec n, \vec n + {\vec \delta}} ~\rangle~,~
\langle~ Y_{\vec n, \vec n + {\vec \delta}} ~\rangle ~$, $~\lambda_n~$
and $~\rho_n~$ are all constants independent
of $~\vec n~$. With this ansatz, the Hamiltonian can be diagonalized in
momentum space by a Bogoliubov transformation (analogous to Eq. (2.18)~)
to obtain
$$\eqalign{{H_{MF} \over N}~=~- ~\lambda S ~&+~ \rho ~X_{\vec 0}^2 ~+ ~{1
\over 2} ~\sum_{\vec \delta} ~J_{\vec \delta}~
(~\vert ~X_{\vec \delta} ~\vert^2 ~-~
\vert ~Y_{\vec \delta} ~\vert^2 ~) ~ \cr
&+ ~\int ~Dk ~(~\omega_{\vec k} ~{\vec A}_{\vec k}^{\dag} \cdot {\vec A}_{
\vec k} ~+ ~{3 \over 2} ~\omega _{\vec k} ~-~
{3 \over 2} ~\mu_{\vec k}~), \cr}
\eqno(3.7)$$
where
$$\eqalign{\mu_{\vec k}~ &=~ \lambda ~+~ \sum_{\vec \delta}~
J_{\vec \delta} ~Y_{\vec \delta} ~\cos {\vec k} \cdot {\vec \delta} \cr
\nu_{\vec k}~ &=~ \rho X_{\vec 0} ~+~ \sum_{\vec \delta}~
J_{\vec \delta} ~(X_{\vec \delta} ~e^{i \vec k \cdot
{\vec \delta}} ~+~ h.c. ~) \cr
{\rm and} {\quad} \omega_{\vec k}~ &=~
(\mu_{\vec k}^2 ~- ~\nu_{\vec k}^2)^{1 \over 2}. \cr}
\eqno(3.8)$$
Unlike the \sw\ spectrum, we now have three decoupled bosons
for each $~{\vec k}$, because the two constraints at each site have been
relaxed
by the MF ansatz to two global constraints.

The variational equations of motion are obtained from the MF \gs \ energy
$$\eqalign{{{E_{MF}} \over N} ~=~ - ~\lambda ~(~S ~+ ~{3 \over 2}~) ~&+~
\rho X_{\vec 0}^2~ + ~{1 \over 2} ~\sum_{\vec \delta} ~J_{\vec \delta} ~(~\vert
{}~
X_{\vec \delta}~ \vert^2 ~-~ \vert ~Y_{\vec \delta} \vert^2 ~) \cr
&+~ {3 \over 2} ~\int~ Dk ~\omega_{\vec k} \cr}
\eqno(3.9)$$
by extremizing it with respect to
$~\lambda, ~\rho, ~X_{\vec \delta}~$ and $ ~Y_{\vec \delta}~$.
We get
$$\eqalign{X_{\vec 0}^2 ~+~ {3 \over 2} ~\int ~Dk~
{{\nu_{\vec k}} \over {\omega_{\vec k}}} ~X_{\vec 0} ~&= ~0, \cr
2~\rho ~X_{\vec 0} ~+~ {3 \over 2} ~\int ~Dk~ {{\nu_{\vec k}}
\over {\omega_{\vec k}}} ~\rho ~&= ~0, \cr}
\eqno(3.10)$$
and
$$\eqalign{{3 \over 2} ~\int~ Dk~ {{\mu_{\vec k}} \over {\omega_{
\vec k}} } ~&=~ S ~+~ {3 \over 2}, \cr
{3 \over 2} ~\int~ Dk~ {{\nu_{\vec k}} \over {\omega_{\vec k}} } ~&=~ 0, \cr
{3 \over 2} ~\int~ Dk~
{{\nu_{\vec k}} \over {\omega_{\vec k}} } ~e^{i {\vec k} \cdot {\vec
\delta}} ~&=~ X_{\vec \delta}, \cr
{3 \over 2} ~\int~ Dk~
{{\mu_{\vec k}} \over {\omega_{\vec k}}} ~\cos {\vec k} \cdot {\vec
\delta} ~&=~Y_{\vec \delta}. \cr}
\eqno(3.11)$$
Eqs. (3.10) imply that $\rho X_{\vec 0}~=~0$.
As a result, $\nu_{\vec k}~$ and $\omega_{\vec k}~$
and, consequently, Eqs. (3.11) do not depend on $~\rho X_{\vec 0}~$. Hence,
we only need to solve Eqs. (3.11).

\vskip .2in

\line {\bf B. ~Condensates and Long-Range Order in Two Dimensions \hfill}

\vskip .1in

Let us first understand what we expect from the \v\ \e\ .  To
leading order in $S$, we expect to recover the \sw\ results,
because the difference between $~X_{\vec \delta}~$ and $~Y_{\vec
\delta}~$ as defined here and in Sec. II, is only that here they
are expectation values of bilinears in \b ic operators, while in
Sec. II, they were products of expectation values of single \b
ic operators.  These two definitions must coincide as the
importance of quantum fluctuations decreases i.e. as $~ S ~\to ~\infty~$.
Hence, in this limit, \sw\ results tell us that
\exi\ , \yi\ , \mi\ , \ni\ and \wi\ are all of $O(S)$.
This means that Eqs. (3.11) cannot be satisfied because the
left-hand sides (LHS) are of $O(1)$, whereas the right-hand
sides (RHS) are of $O(S)$.  (In one dimension, the LHS were at
least singular when \wi\ vanished.  Hence, the \e\ could be
satisfied by allowing for the formation of a gap [~16~]. This is
consistent with the expectation that AFMs are disordered in one
dimension). In two dimensions, the LHS are not singular (except
for special values of $~J_{\vec \delta}~$). Moreover, the
expectation is that AFMs in two or more dimensions are generally
ordered at zero temperature.

The mismatch between the LHS and RHS of Eqs. (3.11) is resolved by
realising that the bosons $A_{\vec k} ~$ condense at the momenta
${\vec k}_i ~, i~=~1, 2, ... ~$, where ${\omega}_{{\vec k}_i} ~$ is
{\it exactly} zero.
This is the signature for an ordered \gs\ in the \b ic language and it is
implemented in the equations by replacing
$${1 \over {2{\omega}_{\vec k}}}~
\to~ {1 \over {2{\omega}_{\vec k} }} ~+~ {{{2\pi}^2} \over v}~
\sum_i ~{\alpha}_i ~{\delta}^2 (~{\vec k} ~-~ {\vec k}_i ~)
\eqno(3.12)$$
on the LHS. (Eq. (3.12) will be justified in the next section
where we study the theory at finite temperatures and then show that
this replacement is appropriate as $T ~\to ~0$). Thus Eqs. (3.11)
are replaced by
$$\eqalign{{3 \over 2}~ \int~ Dk~ {{\mu}_{\vec k}
\over {\omega}_{\vec k}} ~+~ 3 ~\sum_{i}~ {\alpha}_i ~{\mu}_{{\vec k}_i} ~&=~
S ~+~ {3 \over 2}, \cr
{3 \over 2} ~\int ~Dk~ {{\nu}_{\vec k} \over {\omega}_{\vec k}}~
+~ 3 ~\sum_{i} ~{\alpha}_i ~{{\nu}_{\vec k}}_i ~&=~ 0, \cr
{3 \over 2} ~\int~ Dk~ {{\nu}_{\vec k}
\over {\omega}_{\vec k}} ~e^{i \vec k \cdot {\vec \delta}}~
+~3 ~\sum_{i} ~{\alpha}_i ~{{\nu}_{\vec k}}_i ~e^{i {\vec k}_i
\cdot {\vec \delta}} ~&=~ X_{\vec \delta}, \cr {\rm and} \quad
{3 \over 2} ~\int~ Dk~ {{\mu}_{\vec k} \over
{\omega}_{\vec k}} ~{\cos {\vec k} \cdot {\vec \delta}}~
+~ 3 ~\sum_{i}~ {\alpha}_i ~{{\mu}_{{\vec k}_i}~
\cos {{\vec k}_i \cdot {\vec \delta}} } ~&=~ Y_{\vec \delta}. \cr}
\eqno(3.13)$$
To solve these \e\ to leading order in $S$, the integrals which are
of $O(1)$ are dropped, and \exi\ , \yi\ , $\lambda$ and $\rho$ are
substituted from Eqs. (2.14) and (2.15). We then obtain the values of
$~{\alpha}_i~$.

In category (i) defined in Sec. II, the pitch vector $\vec Q$
and $~- ~\vec Q$ do not differ by a reciprocal lattice vector.
We then find that \wi\ vanishes at $\vec k ~=~ 0, ~{\vec Q}~$
and $~ -~ {\vec Q}~$ and hence, there exists condensates at
these values of $\vec k~$. Their amplitudes $\alpha _{\vec 0} ~,
\alpha_{\vec Q}~$ and $~\alpha_{- \vec Q} ~$ are given by
$$\eqalign{{1 \over {3 \alpha_{\vec 0}}} ~&=~ 2 ~\sum_{\vec \delta} ~
J_{\vec \delta} ~ {\sin}^2 (~{\vec Q} \cdot {\vec \delta} / 2 ~)
\cr {\rm and} {\quad} {1 \over {3 \alpha_{\vec Q}}} ~=~ {1
\over {3 \alpha_{- \vec Q}}} ~&=~ - ~4 ~\sum_{\vec \delta}~
J_{\vec \delta} ~{\sin}^2 (~{\vec Q \cdot {\vec \delta}} / 2 ~) ~\cos
{\vec Q} \cdot {\vec \delta}.  \cr}
\eqno(3.14)$$
For category (ii), $2\vec Q~=~ {\vec \tau}$ and there exist
condensates only at $\vec k~=~ \vec 0 ~$ and $~ \vec Q$, with
$~\alpha_{\vec 0}~$ given by Eq. (3.14) and $~\alpha_{\vec Q}~$
given by $${1 \over {3 \alpha_{\vec Q}}} ~=~ -~2 ~\sum_{\vec
\delta} ~J_{\vec \delta}~ {\sin}^2 (~{\vec Q} \cdot {\vec
\delta} / 2 ~) ~\cos {\vec Q} \cdot {\vec \delta}.
\eqno(3.15)$$

The categories (iii) and (iv) are less important because the order from
disorder phenomenon always picks out a \gs\ of the form in (ii). However,
if we work with an arbitrary state in these two categories,
we expect condensates at $~\vec k~=~ \vec 0,~ {\vec Q}_1~,~ {\vec Q}_2~$
and $~ {\vec Q}_3~$.

These condensates are related to LRO in the system. To see this,
note that we can write
$$\langle~ {\vec S}_{\vec 0} \cdot {\vec
S}_{\vec n}~\rangle ~=~ Y^2 _{\vec n} ~-~ X_{\vec n}^2
\eqno(3.16)$$
where
$$\eqalign{Y_{\vec n}~ &=~ \langle~{\vec a}_{\vec 0}^{\dag} \cdot
{\vec a}_{\vec n}~ \rangle \cr
{\rm and} {\quad} X_{\vec n}~ &=~ \langle~{\vec a}_{\vec 0} \cdot
{\vec a}_{\vec n} ~\rangle . \cr}
\eqno(3.17)$$
In terms of Fourier components,
$$Y_{\vec n} ~\pm ~X_{\vec n} ~=~ {3 \over 2} ~\int~ Dk~
\bigl( ~{{{\mu}_{\vec k} ~\pm~ {\nu}_{\vec k}} \over {{\omega}_{\vec k}}}~
\bigr) ~e^{i \vec k \cdot \vec n}~
+ ~3 ~\sum_i~ {\alpha}_i ~( {\mu}_{{\vec k}_i} ~\pm ~{\nu}_{{\vec k}_i}~) ~
e^{i {\vec k}_i \cdot \vec n}.
\eqno(3.18)$$
To $~O(S)~$, this gives
$$\langle~ {\vec S}_{\vec 0}
\cdot {\vec S}_{\vec n}~\rangle ~=~ S^2 ~\cos {\vec Q} \cdot \vec n
\eqno(3.19)$$
as expected for an ordered \gs\ .  Finally, we should point out that the
energy \wi\ can vanish to $O(S)$ for other values of $\vec k$ also.
The values of $\vec k$ where condensates are formed are called
Goldstone modes. Here, \wi\ is {\it exactly} zero for reasons of symmetry
breaking.
At the other values of $\vec k$, the condensate density is zero and  \wi\
is zero {\it only} to $O(S)$ and quantum effects produce a gap at higher
orders in ${1 / S}~$, called the quantum exchange gap. We
refer the reader to Ref. [~7~] for further details.

\vfill
\eject

\line{\bf IV. ~FINITE-TEMPERATURE ANALYSIS \hfill}

\vskip .1in

We now examine what happens at finite but low temperatures. By low
we mean that $~T ~<<~ J S^2~$, where $J$ is a typical coupling
strength. We shall confirm the general expectation that two-dimensional
AFMs are disordered at finite \zte\ by computing the
spin-spin correlation and showing that it decays exponentially
beyond a certain length scale.
We will first justify the zero-\zte\ ansatz Eq. (3.12) used in the
previous section and then derive the form of the correlation length $~\xi~$
at finite $~T$.

At finite \zte\ , the mean-field free energy of the model in Eq. (3.8) is
given by
$${{F_{MF}} \over N} ~=~ {{E_{MF} (T=0)} \over N} ~-~ 3T ~\int ~Dk ~\ln~ (~1 ~
+~ \langle~n_{\vec k} ~\rangle~ )
\eqno(4.1)$$
where $~E_{MF}(T=0)~$  is the \gs\ energy defined in Eq. (3.9), and
$\langle~n_{\vec k}~\rangle~$ is the occupation number given by
$$\langle~n_{\vec k}~\rangle ~=~
(~\exp (~{{\omega}_{\vec k} / T}~) ~-~1~)^{-1}.
\eqno(4.2)$$
(We have set the Boltzmann constant equal to one.) The \v\ \e\ are again
obtained by extremizing $F_{MF}~$ with respect to the parameters
$~\lambda, ~\rho, ~X_{\vec \delta}~, ~Y_{\vec \delta}~$.
If $p$ denotes any one of these parameters, then it is clear that
${1 \over 2} ~\partial {\omega}_{\vec k} / \partial p~$ at
$~T~=~0$ is replaced by $~(~\langle~n_{\vec k}~\rangle ~+~ {1 \over 2}~)~
\partial {\omega}_{\vec k} / \partial p ~$ at finite $~T~$. If \wi\
is non-zero, then $\langle~n_{\vec k}~\rangle ~\to~ 0$ very rapidly as
$~T ~\to ~0~$. But if \wi\ is {\it exactly} zero at $~\vec k ~=~ {\vec k}_i ~$,
then $~\langle~n_{\vec k} ~\rangle~ $ diverges as $\vec k$ approaches
${\vec k}_i~$ for any {\it finite} $~T$.
If $~{\omega}_{\vec k}\ \partial {\omega}_{\vec k} / \partial
p~$ remains finite as $~\vec k ~\to ~{\vec k}_i ~$,  then the neighborhood
of the point ${\vec k}_i~$ leads to an integral like
$$\int_{\vec k \sim {\vec k}_{i}} ~Dk~ {{ \langle~n_{\vec k}~\rangle} \over
{{\omega}_{\vec k}}}
\eqno(4.3)$$
multiplying
$(~{\omega}_{\vec k} ~\partial {\omega}_{\vec k} / \partial p ~)_{\vec
k ~=~ {\vec k}_{i}}~$ in the \v\ \e\ .

In two dimensions, the integral Eq. (4.3) diverges logarithmically
if $~{\omega}_{\vec k}^2 ~\sim~ (\vec k ~-~ {\vec k}_i)^2 ~$
near $~{\vec k}_i~$ (which is most common) and is
even more divergent if $~{\omega}_{\vec k}^2 ~$ goes to zero
faster than quadratically.
For finite $T$, this divergence is cured by the formation of a
small gap at $~{\vec k}_i~$. But as $~T ~\to~ 0~$, the gap
vanishes exponentially and the divergence (which occurs in a
region which shrinks to zero with $T$) is a signal that the
bosons want to condense at ${\vec k}_i~$. At $~T~=~0~$, this
information is fed into Eq. (4.3) by the replacement
$${{\langle~n_{\vec k}~\rangle}
\over {{\omega}_{\vec k}}} ~\to~ {{(2\pi)^2} \over v}~
{\alpha}_i ~{\delta}^2 (~\vec k ~-~ {\vec k}_i~)
\eqno(4.4)$$
near $~\vec k ~=~ {\vec k}_i~$ where $~\alpha_i~$ is the amplitude
for the \b s to condense at $~{\vec k}_i~$. This argument
justifies Eq. (3.12).

At a Goldstone mode ${\vec k}_i~$, the square of the energy is
related to the components $(\vec k ~- ~{\vec k}_i)_{\sigma}~$, (where
$~\sigma ~=~ x, y ~$), as
$${\omega}_{\vec k}^2 ~=~ \sum_{\sigma ,\eta} ~V_{\sigma \eta} ~(~\vec k ~-~
{\vec k}_i~)_{\sigma} ~(~\vec k ~-~ {\vec k}_i~)_{\eta}
\eqno(4.5)$$
where $V$ is the velocity matrix.
We can diagonalize the real symmetric matrix $V$ by an
orthogonal transformation to obtain its eigenvalues
$~C_{xi}^2 ~$ and $ ~C_{yi}^2~$ which are the
\sw\ velocities at $~{\vec k}_i~$.
We shall assume that the $~C_{\sigma i}~$
are finite,~-~ $~i.e.$,
$~{\omega}_{\vec k}^2~$ vanishes only quadratically as
$~\vec k ~\to~ {\vec k}_i~$ along any direction. These velocities will be
of $O(S)$ for large $S$.

For finite $T$, there is no LRO in $d~=~2$ and therefore no condensates.
Instead we have {\it gaps} $~\Delta_i~$ at all the Goldstone modes.
Just as Eqs. (3.11) are dominated at $O(S)$ by the condensates,
the low-temperature \e\ will be dominated at $O(S)$ by the neighborhoods
of $~{\vec k}_i~$. We make the ansatz that at
low $T$, $~{\omega}_{\vec k}^2~$ near
$~{\vec k}_i~$ simply differs from Eq. (4.5) by a small gap
$~\Delta_i~$. (We shall soon verify the self-consistency of this ansatz). Thus,
$${{\omega}_{\vec k}}^2 ~=~ {\Delta}_{i}^{2} ~+~ C_{xi}^{2} ~{\delta
{k_x}}^{2}~ +~ C_{yi}^{2} ~{\delta {k_y}}^2
\eqno(4.6)$$
where the $~\delta k_{\sigma}~$ are related to $~(~\vec k ~-~ {\vec k}_i~)_{
\sigma}~$ by a rotation. We then integrate both sides of Eq. (4.4) over a
neighborhood of $~{\vec k}_i~$ where
$\langle~n_{\vec k}~\rangle ~\simeq ~T / {\omega}_{\vec k} ~$ to
find a relation between the $T~=~0$ condensate
$~\alpha_i~$ and the low-temperature gap $~\Delta_i~$ given by
$${\Delta}_i~ \sim~ {\bigl(~ {{C_{xi}~ C_{yi}} \over v} ~\bigr)}^{1 \over 2}
\exp~ [~ -~ {{2\pi ~C_{xi}~ C_{yi} ~{\alpha}_{i}} \over {v ~T}}~].
\eqno(4.7)$$
So the gaps $~{\Delta}_{i}~ \sim~ \exp (~-~ \pi J S^2 / T~)~$ are
indeed exponentially small for small $T$.

Finally, we examine how the spin-spin correlation in Eqs. (3.18) and (3.19)
is modified at low temperature. For $~\vec n \ne \vec 0~$, the expressions for
$~X_{\vec n}~$ and $~Y_{\vec n}~$ are now
$$Y_{\vec n} ~\pm ~X_{\vec n} ~=~
3 ~\int ~Dk~ \bigl(~{{{\mu}_{\vec k} ~\pm~ {\nu}_{\vec k}} \over {{\omega}_{
\vec k}}} ~\bigr)~
(~\langle~n_{\vec k}~\rangle ~+~ {1 \over 2}~) ~e^{i \vec k \cdot \vec n}.
\eqno(4.8)$$
At low temperatures, this is dominated by the regions near the
Goldstone points.  We then find that the correlation falls off
exponentially with the distance $~\vert ~{\vec n} ~\vert~$. To
take a simple and specific example, consider categories (i) and
(ii) and assume isotropy of velocities $~C_{xi} ~=~ C_{yi} ~=~
C_i~$ at $~\vec 0~$ and $~\pm ~\vec Q~$.  The correlation is
then given by
$$\langle~{\vec S}_{\vec 0} \cdot {\vec
S}_{\vec n}~\rangle~
\sim {{T^2} \over {{J^2} {S^2}}} ~~\cos {\vec Q} \cdot {\vec n}~~
e^{-~ {\vert \vec n \vert} / \xi}
\eqno(4.9)$$
for $~\vert~ \vec n ~\vert >> ~\xi~$, where
$${1 \over {\xi}}~=~ {{{\Delta}_{\vec 0}} \over {C_{\vec 0}}} ~+~
{{{\Delta}_{\vec Q}} \over {C_{\vec Q}}}.
\eqno(4.10)$$

\vskip .2in

\line{\bf V. ~DISCUSSION \hfill}

\vskip .1in

We have used a three boson representation for spins in order to
develop a Hartree-Fock MFT for \fr\ \ha\ AFMs in two dimensions.
At zero temperature, the MF \v\ \e\ are satisfied by forming
condensates of the bosons at the Goldstone points which implies
LRO. At finite \zte s, the condensates and LRO disappear, to be
replaced by gaps at the Goldstone points and exponential
fall-off of the spin correlations.

An interesting feature of this MFT is its connection to a new
\sw\ theory which we have introduced in this paper. This new theory
is applicable when the spin ordering is fully three-dimensional.
Another interesting feature of our formalism is that it can also
be used to address the question of {\it disordered}  \gs s. For
instance, in certain regions of the space of couplings (e.g.
$~J_{2}~=~ J_1 / 2~$ on the square lattice), the question of
whether the \gs\ is ordered or disordered even at $T=0$ is
still open. Within our formalism, this would depend on whether
the energy \wi\ vanishes faster than quadratically as $\vec k$
approaches a Goldstone point.  If that happens, the integrals
would diverge and the \v\ \e\ would be satisfied, not by
condensates, but by gaps in the spectrum. Such gaps could imply
the absence of LRO and, perhaps, spin-liquid states. This
formalism can also be used to investigate the possibility of \gs
s which break translational symmetry (e.g. dimerized states) by
allowing the \v\ parameters to have different periodicities.

However the 3BR has a disadvantage as well. The analytic
expressions for the gaps differ by an overall factor of three
from similar expressions derived from other methods [~7,~18~].
This disagreement stems from the basic problem that the 3BR
overcounts the number of \sw\ modes by a factor of three. One
must therefore be cautious when comparing numbers from our
calculations with actual numerical simulations of the \ha\ AFM.

\vskip .2in

We thank D. M. Gaitonde and H. R. Krishnamurthy for useful discussions.

\vfill
\eject

\line{\bf References \hfill}

\vskip .2in

\item{1.}{P. W. Anderson, Science, {\bf 235}, 1196 (1987).}

\item{2.}{S. Chakravarty, B. I. Halperin and D. R. Nelson, Phys. Rev. B
{\bf 39}, 2344 (1989).}

\item{3.}{T. Dombre and N. Read, Phys. Rev. B {\bf 38}, 7181 (1988); E.
Fradkin and M. Stone, {\it ibid.} {\bf 38}, 7215 (1988); X. G. Wen and A.
Zee, Phys. Rev. Lett. {\bf 61}, 1025 (1988); F. D. M. Haldane, {\it ibid.}
{\bf 61}, 1029 (1988).}

\item{4.}{L. B. Ioffe and A. I. Larkin, Int. J. Mod. Phys. B {\bf 2}, 203
(1988).}

\item{5.}{P. Chandra and B. Doucot, Phys. Rev. B {\bf 38}, 9335 (1988).}

\item{6.}{N. Read and S. Sachdev, Phys. Rev. B {\bf 42}, 4568 (1990) and
Phys. Rev. Lett. {\bf 62}, 1694 (1989).}

\item{7.}{P. Chandra, P. Coleman and A. I. Larkin, J. Phys. Cond. Matt.
{\bf 2}, 7933 (1990).}

\item{8.}{I. Affleck and J. B. Marston, Phys. Rev. B {\bf 37}, 3744 (1988);
J. B. Marston and I. Affleck, {\it ibid.} {\bf 39}, 11538 (1989).}

\item{9.}{X. G. Wen, F. Wilczek and A. Zee, Phys. Rev. B {\bf 39}, 11413
(1988).}

\item{10.}{D. P. Arovas and A. Auerbach, Phys. Rev. B {\bf 38}, 316 (1988);
A. Auerbach and D. P. Arovas, Phys. Rev. Lett. {\bf 61}, 617 (1988).}

\item{11.}{S. Sarkar, C. Jayaprakash, H. R. Krishnamurthy and M. Ma, Phys.
Rev. B {\bf 40}, 5028 (1988); D. Yoshioka, J. Phys. Soc. Jpn. {\bf 58}, 32
(1989).}

\item{12.}{F. Mila, D. Poilblanc and C. Bruder, Phys. Rev. B {\bf 43}, 7891
(1991).}

\item{13.}{J. Ferrer, Rutgers preprint (1992) (unpublished).}

\item{14.}{A. Moreo, E. Dagotto, Th. Jolicoeur and J. Riera, Phys. Rev. B
{\bf 42}, 6283 (1990).}

\item{15.}{J. D. Reger and A. P. Young, Phys. Rev. B {\bf 37}, 5978 (1988).}

\item{16.}{S. Rao and D. Sen, IISc preprint IISc-CTS-92-12 (1992), to appear
in Phys. Rev. B.}

\item{17.}{J. Villain, J. Phys. (Paris) {\bf 38}, 385 (1977).}

\item{18.}{A. V. Chubukov and Th. Jolicoeur, Phys. Rev. B {\bf 46}, 11137
(1992); Th. Jolicoeur, E. Dagotto, E. Gagliano and S. Bacci, Phys. Rev. B
{\bf 42} 4800 (1990).}

\vfill
\eject

\line{\bf Table Captions \hfill}

\vskip .2in

\item{[1]}{Categories of classical ground states on the square and triangular
lattices for various ratios of the coupling constants $~J_1 ~$ and $~J_2 ~$.}

\item{[2]}{Table of the angles $~\phi_{\vec \delta} ~$ and the short-range
order parameters $~X_{\vec \delta} ~$ and
$~Y_{\vec \delta} ~$ as a function of $~{\vec \delta} ~$ for category (iv) ---
the three-dimensional canted state.}

\vfill
\eject

\vbox{\tabskip=0pt \offinterlineskip
\def\tablerule{\noalign{\hrule}}
\halign to400pt{\strut#& \vrule#\tabskip=1em plus2em&
\hfil#& \vrule#& \hfil#& \vrule#&
\hfil#& \vrule#\tabskip=0pt \cr
\tablerule
&&\omit&&\omit&&\omit& \cr
&&\omit\hidewidth Lattice \hidewidth&&
\omit\hidewidth Coupling Strengths \hidewidth&&
\omit\hidewidth Category of  \hidewidth& \cr
&&\omit&&\omit&&\omit\hidewidth Ground State \hidewidth& \cr
&&\omit&&\omit&&\omit& \cr
\tablerule
&&\omit&&\omit&&\omit& \cr
&&\hidewidth Square \hidewidth&&\hfill$J_2 ~<~ J_1 /2 $\hfill&& (ii) & \cr
&&\omit&&\omit&&\omit& \cr
\tablerule
&&\omit&&\omit&&\omit& \cr
&& &&\hfill$J_2 ~ =~ J_1 /2$\hfill&& (iv) & \cr
&&\omit&&\omit&&\omit& \cr
\tablerule
&&\omit&&\omit&&\omit& \cr
&& &&\hfill$J_2 ~> ~ J_1 /2$\hfill&& (iii) & \cr
&&\omit&&\omit&&\omit& \cr
\tablerule
&&\omit&&\omit&&\omit& \cr
\tablerule
&&\omit&&\omit&&\omit& \cr
&& Triangular &&\hfill$J_2 ~<~ J_1 /8 $\hfill&& (i) & \cr
&&\omit&&\omit&&\omit& \cr
\tablerule
&&\omit&&\omit&&\omit& \cr
&& &&\hfill$J_1 /8 ~\le ~J_2 ~\le~ J_1 $\hfill&& (iv) & \cr
&&\omit&&\omit&&\omit& \cr
\tablerule
&&\omit&&\omit&&\omit& \cr
&& &&\hfill$J_2 ~> ~ J_1 $\hfill&& (i) & \cr
&&\omit&&\omit&&\omit& \cr
\tablerule}}

\vskip .4in

\centerline{\bf Table 1}

\vfill
\eject

\vbox{\tabskip=0pt \offinterlineskip
\def\tablerule{\noalign{\hrule}}
\halign to450pt{\strut#& \vrule#\tabskip=1em plus2em&
\hfil#& \vrule#& \hfil#& \vrule#& \hfil#& \vrule#&
\hfil#& \vrule#& \hfil#& \vrule#\tabskip=0pt \cr
\tablerule
&&\omit&&\omit&&\omit&&\omit&&\omit& \cr
&&\omit\hidewidth$\cos {\vec Q_1} \cdot {\vec \delta}$\hidewidth&&
\omit\hidewidth$\cos {\vec Q_2} \cdot {\vec \delta}$\hidewidth&&
\omit\hidewidth $~\cos \phi_{\vec \delta}~$ \hidewidth&&
\omit\hidewidth $~X_{\vec \delta}~ / ~S ~$ \hidewidth&&
\omit\hidewidth $~Y_{\vec \delta}~ / ~S ~$\hidewidth& \cr
&&\omit&&\omit&&\omit&&\omit&&\omit& \cr
\tablerule
&&\omit&&\omit&&\omit&&\omit&&\omit& \cr
&&\hfill$1$\hfill&&\hfill$1$\hfill&&\hfill$1$\hfill&&\hfill
$0$\hfill&&\hfill$1$\hfill& \cr
&&\omit&&\omit&&\omit&&\omit&&\omit& \cr
\tablerule
&&\omit&&\omit&&\omit&&\omit&&\omit& \cr
&&\hfill$-1$\hfill &&\hfill$-1$\hfill && \hfill$\cos 2 \theta $\hfill &&
\hfill$\sin^2 \theta$\hfill
&& \hfill$-\cos^2 \theta$\hfill & \cr
&&\omit&&\omit&&\omit&&\omit&&\omit& \cr
\tablerule
&&\omit&&\omit&&\omit&&\omit&&\omit& \cr
&& \hfill$1$\hfill && \hfill$-1$\hfill && \hfill$\sin^2 \theta ~\cos 2 \phi $
\hfill && \hfill
$\sin^2 \phi - \cos^2 \theta ~\cos^2 \phi $\hfill && \hfill$ \sin^2 \theta ~
\cos^2 \phi $\hfill & \cr
&&\omit&&\omit&&\omit&&\omit&&\omit& \cr
&&\omit&&\omit&& \hfill$- \cos^2 \theta $\hfill && \hfill$+
i ~\cos \theta ~\sin 2 \phi $ \hfill&&\omit& \cr
&&\omit&&\omit&&\omit&&\omit&&\omit& \cr
\tablerule
&&\omit&&\omit&&\omit&&\omit&&\omit& \cr
&& \hfill$-1$\hfill && \hfill$1$\hfill &&
\hfill$ - \sin^2 \theta ~\cos 2 \phi $\hfill &&
\hfill$\cos^2 \phi - \cos^2 \theta ~\sin^2 \phi $\hfill &&
\hfill$ \sin^2 \theta ~\sin^2 \phi$\hfill& \cr
&&\omit&&\omit&&\omit&&\omit&&\omit& \cr
&&\omit&&\omit&& \hfill$- \cos^2 \theta $\hfill &&
\hfill$- i ~\cos \theta ~\sin 2\phi $\hfill &&\omit& \cr
&&\omit&&\omit&&\omit&&\omit&&\omit& \cr
\tablerule}}

\vskip .4in

\centerline{\bf Table 2}

\vfill
\eject

\line{\bf Figure Caption \hfill}

\vskip .2in

\item{[1]}{Brillouin zone for the triangular lattice. The points A, B and
C denote the pitch vectors of the classical ground states for $~
\epsilon ~=~ J_2 / J_1 ~$ lying in the
ranges $~\epsilon ~<~ 1/8 ~$, $~1/8 ~\le ~\epsilon ~\le ~1 ~$ and $~
\epsilon~ > ~1$  respectively.}

\end